\documentclass[prd, onecolumn, nofootinbib, floatfix, notitlepage]{revtex4-2}

\usepackage{amsmath}
\usepackage{graphicx}
\usepackage{dcolumn}
\usepackage{bm}
\usepackage{epsfig}
\usepackage{amssymb,latexsym,mathrsfs}
\usepackage{graphicx}
\usepackage{color}
\usepackage{xcolor} 
\usepackage{mathtools} 
\usepackage{hyperref}

\hypersetup{
    colorlinks=true,
    linkcolor=red,
    citecolor=blue,
} 

\newcommand{\be}{\begin{equation}}
\newcommand{\ee}{\end{equation}}
\newcommand{\bs}{\begin{split}} 
\newcommand{\bea}{\begin{eqnarray}}
\newcommand{\eea}{\end{eqnarray}}

\newcommand{\om}{\Omega_m} 
\newcommand{\oma}{\Omega_{m,1}} 
\newcommand{\omb}{\Omega_{m,2}} 
\newcommand{\lcdm}{$\Lambda$CDM}

\begin{document}

\title{A Whole Cosmology View of the Hubble Constant} 

\author{Eric V. Linder${}^{1,2}$} 
\affiliation{
${}^1$Berkeley Center for Cosmological Physics \& Berkeley Lab, 
University of California, Berkeley, CA 94720, USA\\ 
${}^2$Energetic Cosmos Laboratory, Nazarbayev University, Astana, Qazaqstan 010000\\ 
} 

\begin{abstract}
The Hubble constant $H_0$ is the value of the cosmic expansion 
rate at one time (the present), and cannot be adjusted successfully without 
taking into account the entire expansion history and 
cosmology. We outline some conditions, that if not quite 
``no go'' are ``no thanks'', showing that changing the expansion 
history, e.g.\ employing dynamical dark energy, cannot reconcile 
disparate deductions of $H_0$ without upsetting some other 
cosmological measurement. 
\end{abstract} 

\date{\today} 

\maketitle

\section{Introduction} 

Cosmology is the study of the universe as a whole. This 
is usually interpreted as {\it Cosmology is the study 
of the (universe as a whole)\/}, where universe as a 
whole means large spatial scales and large volumes. 
However one should also be aware that {\it Cosmology is 
the study of the universe (as a whole)\/}, where as a 
whole means both over a wide range of history and over 
the full panoply of observations. This has two essential 
implications for the Hubble constant discussion: 

\begin{enumerate} 

\item $H_0\equiv H(z=0)$ is merely a value at one instant 
of a dynamical description of cosmic expansion, and it is 
very little use getting it ``right'' without accounting 
properly for the whole dynamics. 

\item $H_0$ is merely one number from the foundational 
functions, e.g.\ cosmic expansion history, cosmic growth 
history, cosmic gravity history, that describe cosmology, 
and cannot be examined in isolation. 

\end{enumerate} 

This chapter focuses on these two aspects of the Hubble 
constant issue, specifically whether changing the 
expansion history through using a dynamical (dark energy) 
density can reconcile different deductions of $H_0$. 

We will give arguments that are as model independent as 
possible, and hence not list the quite numerous models 
that have been considered (see, for example, 
\cite{buyer,olympic,realm,intertwine,hunter} and references therein 
for many examples). Rather we will seek to come as 
close to ``no go'' strictures as possible, though in the 
vast arena of trade offs between standard and created 
model parameters these are closer to ``no thanks''. 
The philosophy follows that of Occam, not to multiply 
entities unless necessary. 

\begin{figure} 
\centering 
\includegraphics[width=0.6\columnwidth]{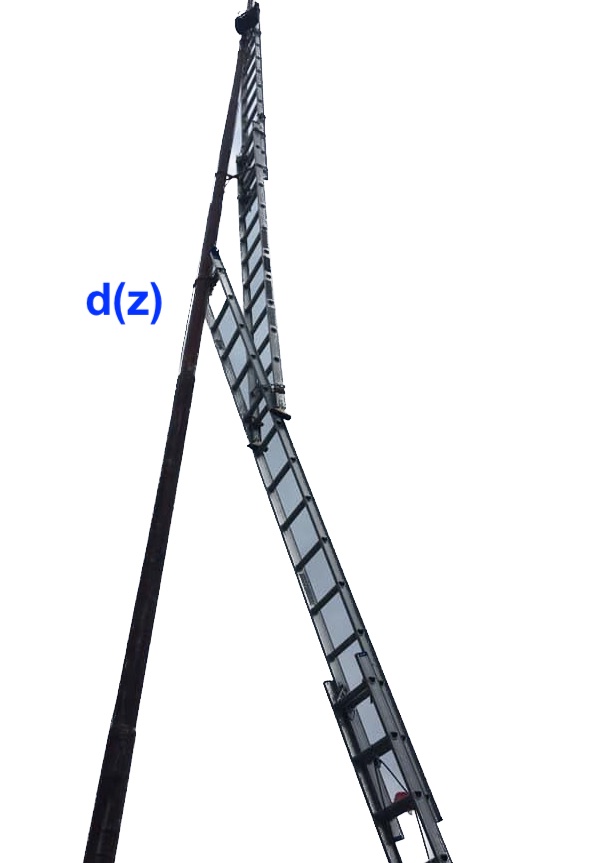}
\caption{Ladders require care in use and 
are most stable when anchored firmly at both ends.}
\label{fig:ladder} 
\end{figure}

One of the simplest 
reconcilings of the Hubble constant values is that some 
measurement is incorrect. Indeed, in the nearly 100 year 
history of the Hubble constant this has always been the 
answer, with much of that history having discrepancies 
at nominal statistical $\sigma$ levels in excess of the 
present situation. We will say no more about this 
resolution beyond Figure~\ref{fig:ladder}, instead 
assuming that the discrepant deductions are all valid and 
seeking for a cosmological paradigm to reconcile them.

\section{Expansion History} \label{sec:exp} 

It is relatively easy to shift the value of $H_0$ 
by changing the cosmological model, but again $H_0$ 
is not the whole of cosmology. To repeat: cosmology 
is the study of the universe as a whole, or alternately, 
all cosmology all the time \cite{allcos,2112.11567}. We need to 
take into account the effects on the full expansion 
history $H(z)$, not just $H(z=0)$. 

Let us being by saying that the cold dark matter 
plus cosmological constant, i.e.\ \lcdm, cosmology 
is a good description of the expansion history. 
That is, we would like to match some expansion 
history model to observations that are adequately 
described within the (flat) cosmology 
\be 
H^2(z)=H^2_2\omb (1+z)^3+H^2_2(1-\omb)\,. 
\ee 
Here $H_2=H(z=0)$ is the Hubble constant for this 
cosmology and we specify that the model has a 
matter density fraction $\omb$. However, when making 
local universe measurements we find the Hubble constant 
is $H_1$ and try to reconcile the situation by 
allowing deviations from \lcdm\ in the form of 
dynamical dark energy. So we take a model 
\be 
H^2(z)=H^2_1\oma (1+z)^3+\frac{8\pi G}{3}\rho_{\rm de}(z)\ . 
\ee 

We only relate these models over a finite range 
of redshift, where observations indicate the \lcdm\ 
model matches the observations well. (One could add 
spatial curvature as well but this does not qualitatively 
change the results.) Thus the necessary dark 
energy density is 
\be 
\frac{8\pi G}{3}\rho_{\rm de}(z)=(H^2_2\omb-H^2_1\oma)(1+z)^3+H^2_2(1-\omb)\ . \label{eq:rhoexp} 
\ee 

We can immediately see why it is so crucial to 
do cosmology as a whole, rather than only think about 
the Hubble constant. If we have observations determining 
$\om h^2$ (the cosmic microwave background (CMB) value for 
what should be a universal constant $\om h^2$ should be 
much less model dependent than the CMB value for $h$ itself) 
then we are forced to conclude that our 
attempt to change the Hubble constant through dynamical 
dark energy fails: one must have $H_2=H_1$. It is only 
when we ignore the rest of cosmology that we can 
obtain a different Hubble constant. 

Suppose that we do not have $\om h^2$ determined by 
the CMB. In that case then Eq.~\eqref{eq:rhoexp} tells 
us the form of dark energy that can shift the Hubble 
constant -- it looks like the sum of a cosmological 
constant and a matter-like term. In particular, if 
the value $H_2$ coming from the \lcdm\ fit to the 
observations (e.g.\ the ``CMB'' value) is less than 
the value $H_1$ coming from the local measurements 
then the effective energy density will generally be 
negative (one has to compensate for the higher 
expansion rate). In any case, dark energy looking 
like the sum of matter and $\Lambda$ has been studied 
for decades, for example in mocker and polytropic 
dark energy \cite{freese02,0601052}; very generically 
it causes growth of structure to deviate significantly 
from the \lcdm\ cosmology. Once again, one must use 
cosmology as a whole when assessing the viability of 
shifting the Hubble constant.

\section{Distance-redshift relation} \label{sec:dist} 

One could attempt to sidestep the ``no thanks'' scenarios 
of the previous section by saying we do not know the 
expansion history that well, and are willing to allow 
$H(z)$ to deviate significantly from the \lcdm\ form. 
This is tenable in 2022, but will shortly be severely 
constrained by radial baryon acoustic oscillation (BAO) 
measurements from the Dark Energy Spectroscopic 
Instrument \cite{desisci,desiinst}. For 
this approach, rather than matching 
$H(z)$ over some redshift range one only matches the 
integral quantity of the distance-redshift relation, 
\be 
d(z)=\int_0^z \frac{dz'}{H(z')}\ . 
\ee 
One can have $H(z)$ dip below the \lcdm\ expectation 
at some redshifts and rise above it at others to 
obtain $d(z)$ near \lcdm\ over some finite redshift range. 

Since distances have been measured fairly accurately, 
and since a deviation in $H(z')$ at some redshift affects 
all $d(z>z')$, this is not that easy to accomplish, 
especially when taking into account CMB distances as well. 
Furthermore, changing $H(z)$ also affects growth of 
structure and generally the large swings needed to 
alter the Hubble constant throw off the growth history 
(including the large scale integrated Sachs-Wolfe 
effect in the CMB, see e.g.\ \cite{1305.4530}). In fact, growth of structure is 
not only tightly bound to the expansion history within 
general relativity but nearly determined by the 
well measured distance to CMB last scattering. 

\begin{figure} 
\centering 
\includegraphics[width=0.8\columnwidth]{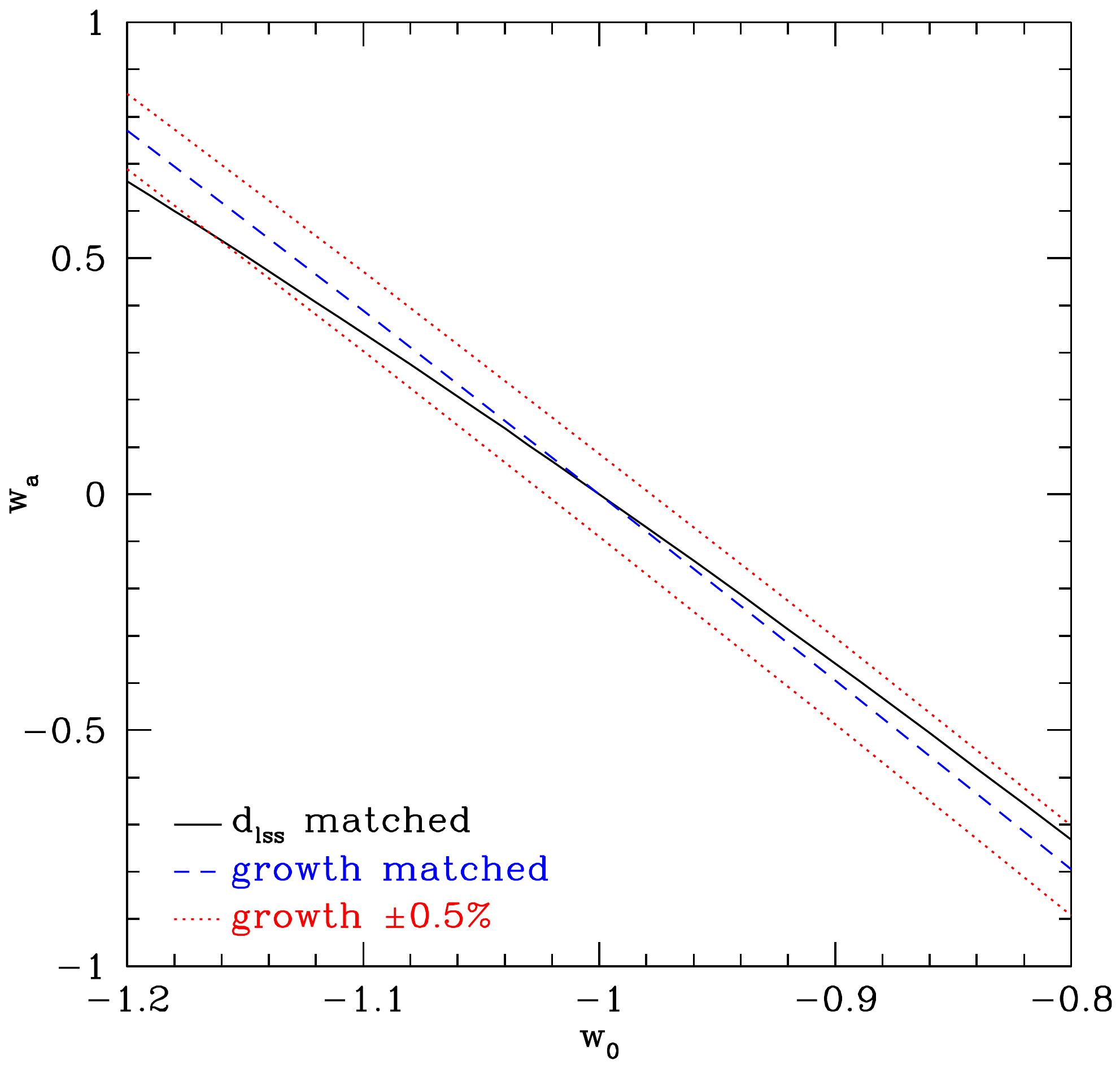}
\caption{Constraints on the distance to CMB last 
scattering $d_{\rm lss}$ tightly constrain structure 
growth as well, even for highly dynamical matching dark energy 
cosmologies. [From \cite{mirage}]}
\label{fig:growcmb} 
\end{figure}

Figure~\ref{fig:growcmb} shows how tightly the 
growth of structure is bound to the distance to 
CMB last scattering, not just for \lcdm\ but for 
quite disparate models along the 
dynamical ``mirage of $\Lambda$'' line. 
Conversely, at low redshift a cartoon relation is 
that the growth rate 
\be 
f\approx\om^{0.55}\sim\left(\frac{\om h^2}{h^2}\right)^{0.55}\ ; 
\ee 
so for preserved $\om h^2$ one has $f\sim h^{-1.1}$, or 
if $f\sigma_8$ is measured then 
$\sigma_8\sim h^{1.1}(f\sigma_8)$.

Thus there is a difficult balancing act in changing 
the Hubble constant, due to the impact of the altered 
expansion history on  growth probes (and the 
CMB through the Sachs-Wolfe effect), 
as well as calibration of supernova distances 
to higher redshift (while preserving 
CMB and BAO distances). All these 
elements must be considered together. 
See \cite{2002.11707,2101.08641,2103.04045,2103.08723,2206.08440} 
for some specific analyses.

\section{Early Universe Deviations} \label{sec:ede} 

One might think of moving alterations of the 
expansion history to high redshift, above where 
distance-redshift constraints apply. This still affects 
growth however, and generally more severely because of 
the long lever arm. One way to get partly around this 
is to push modifications to before the recombination 
epoch. Such an approach is often called early dark energy. 

\begin{figure}
\centering 
\includegraphics[width=0.8\columnwidth]{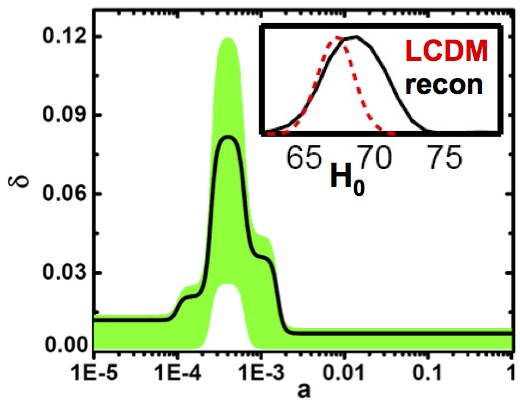}
\caption{Model independent reconstruction of the 
expansion history from CMB data in 2013. The expansion deviation 
$\delta\equiv \delta H^2/H^2_{\Lambda{\rm CDM}}$ shows a 
preference for precombination early dark energy. This is turn 
shifts the Hubble constant $H_0$ from the value 
derived assuming \lcdm\ -- but only slightly, even for 8\% 
early dark energy. [From \cite{hojj}]}
\label{fig:ede} 
\end{figure}

In fact, early dark energy was ``discovered'' in CMB data 
in 2013 \cite{hojj}, along with its concomitant 
shift in the Hubble constant, as shown in 
Figure~\ref{fig:ede}. However the shift was 
much smaller than the current discrepancy, and increasing 
the dark energy density -- raising the early expansion rate 
(and secondarily decreasing the sound horizon) yet 
reducing the clustering matter density fraction -- would have a 
severe effect on both the matter growth history and 
the CMB perturbation power spectrum. Indeed, the effect 
of early dark energy on the sound horizon has been 
familiar for decades 
\cite{efbond,rs2004,doran}, 
but the impact of 
the reduced perturbation potentials is more deleterious, 
forcing a disfavored greater amplitude of mass perturbations 
($\sigma_8$) to compensate and obtain consistency with power spectrum observations. 
It is extremely difficult to obtain 
$H_0>70$ km/s/Mpc and viable CMB and 
viable distances and viable growth. 
See \cite{2103.04999,2010.04158,2003.07355} for some specific examples regarding this. 
Models that push even further back, changing the 
primordial power spectrum beyond the power form, for 
example with oscillations or features, also 
have difficulty in substantially shifting $H_0$, 
e.g.\ \cite{1910.05670,2202.14028} (and will be 
tightly constrained by tomographic redshift surveys 
and future CMB experiments, e.g.\ \cite{2001.07738}).

For postrecombination alterations of the expansion history, 
one again must emphasize the importance of accounting for 
the impact on growth, including the Sachs-Wolfe effect.  
Recall that the Poisson equation imposes 
\bea 
(k/a)^2\Phi&=&4\pi\delta\rho\\ 
&\Rightarrow& \Phi\sim a^2\delta\rho\sim a^2\frac{\delta\rho}{\rho}\rho\\ 
&\Rightarrow& \Phi\sim (a^3\rho)\left(a^{-1}\,\frac{\delta\rho}{\rho}\right)\ . 
\eea 
In the standard scenario the gravitational potential 
is thus time independent and no large Sachs-Wolfe effect 
is generated. However, 
if we change the energy density of clustering matter, 
e.g.\ through adding an 
unclustering early dark energy that influences the recombination 
era or through some interaction, then we affect the growth of 
structure $\delta\rho/\rho\sim a$ (and possibly 
$\rho\sim a^{-3}$) and give a time dependence 
to the gravitational potential $\Phi$, resulting in a large 
Sachs-Wolfe effect contrary to observations. More specifically, 
$\delta\rho/\rho\sim a^{1-(3/5)\Omega_e}$ so only a 
very small fraction of early dark energy $\Omega_e$ 
can be tolerated, hence enabling only a small shift in the 
Hubble constant. 

A rough rule of thumb is that 
\be 
\frac{\delta H_0}{H_0}\sim \frac{1}{2}\,\frac{\delta H^2}{H^2} 
\sim {\Omega_e}\ , 
\ee 
(naively this would be $\sim\Omega_e/2$ but the shifting 
of other parameters to fit the CMB power spectrum 
pushes it closer to $\Omega_e$ \cite{robbers}) 
so a change in $H_0$ to reconcile a 10\% discrepancy in 
values would require $\Omega_e\sim0.1$ and cause a 
$10^{3\times(3/5)\Omega_e}\sim 50\%$ change in the 
growth amplitude today (modulo other shifts). This gives a heuristic explanation 
for why it is nearly impossible to shift $H_0$ from, say,  
$\sim67$ km/s/Mpc to above 70 
(see, for example, \cite{div70,intertwine}) 
and still satisfy distance, 
growth, and CMB measurements -- i.e.\ cosmology as a whole.

\section{Extensions} \label{sec:extend} 

Within the framework used here -- matter plus (an effective) 
dark energy with background and perturbation equations 
governed by general relativity in a 
homogeneous and isotropic universe -- 
reconciliation of 
significantly differing values of the Hubble constant are, 
if not quite ``no go'', then substantially ``no thanks''. 
Due to the impacts on the many other cosmological characteristics, 
mechanisms for shifting $H_0$ cause other alterations away from 
the observed quantities. One could add further physics to 
compensate, e.g.\ interactions, changes in initial conditions, 
etc., but these in turn change a variety of 
observables and 
this runs the risk of becoming epicyclic. Basically 
it requires multiplying entities beyond necessity -- unless 
one's necessity is to obtain a certain value of $H_0$ at 
any cost. Such eschewing of Occam's Razor we regard as 
``no thanks''. 

Furthermore, models that seek reconciliation through a 
mechanism taking place just before recombination or just 
before the present add new fine tuning issues, not simply 
in the model parameters but in why then questions. 
If one seeks a dynamical solution ideally it would arise 
naturally out of the cosmology history. 

Gravity modified from general relativity offers one 
avenue of pursuit: it could compensate for the impact of 
the altered expansion history on the growth of structure, 
or change light propagation in a redshift dependent manner 
to better align the values of the Hubble constant derived 
from strong lens time delay systems. Modified gravity that achieves a  
whole cosmology perspective -- simultaneously providing 
a basis for the current accelerated expansion, growth 
history measurements, and reconciliation of Hubble constant 
values -- would be a theory to explore in depth. 
However it is difficult to see how such a theory would work: 
we generally want gravity to restore to general relativity 
at early times such as CMB recombination and on small 
scales such as the Cepheid distance ladder measurements 
so there should be no shift between the 
$H_0$ values.

\section{Conclusions} 

The history of the Hubble constant across a century 
has always been one of disparate values. In all cases 
those values have shifted as measurement and modeling 
uncertainties have become better studied. There seems 
little reason to expect the current situation to be 
drastically different. 

If a compelling, succinct physical explanation for 
differing values when derived from distinct probes 
awaited, then perhaps one might give more weight to 
new physics relative to systematics. However as we 
have reviewed here, it is remarkably difficult to 
shift the Hubble constant from a given value, even 
in the presence of quite dramatic dynamics at any 
point in cosmic history -- {\it when taking into 
account cosmology as a whole\/}. While this does 
not rise to the level of a no go theorem, it is 
definite enough to weigh heavily as a no thanks 
likelihood. 

A desideratum for a model with new physics would be 
one that gives an advantageous likelihood for the 
full panoply of cosmic observations -- distances, 
growth, and CMB -- without putting a thumb on the 
scale by imposing a prior on the Hubble constant, 
the very issue one is examining. That is postdiction 
not prediction. Plainly put, $H_0$ 
data or prior should not be input into the analysis 
whose aim is examine the value of $H_0$. If the 
posterior for distances plus growth plus CMB observations 
over cosmic history is not advantaged over a 
standard \lcdm\ model, then it does not matter 
what value of $H_0$ can be achieved with the model (the same 
way that if one was trying to reconcile the matter 
density $\om$ then one would not put a prior on that). 

If an alternate model does succeed in fitting 
all cosmology, all the time, then one proceeds 
with Occam's Razor and physical naturalness examinations. 
In the end, more data -- with more understanding of 
the subtleties, selections, systematics 
(see for example \cite{2106.09400,2206.08160} for steps 
in this direction) -- from 
more probes will be the path to resolution, and a 
firm foundation for cosmology as a study of the 
universe as a whole.

\end{document}